*Review*

# COVID-19 Contact-tracing Apps: a Survey on the Global Deployment and Challenges

**Jinfeng Li** [1]* **and Xinyi Guo**

[1] Department of Electrical and Electronic Engineering, Imperial College London; jinfeng.li@imperial.ac.uk
* Correspondence: jinfeng.li@imperial.ac.uk

**Abstract:** To address the massive spike in uncertainties triggered by the coronavirus disease (COVID-19), there is an ever-increasing number of national governments that are rolling out contact-tracing Apps to aid the containment of the virus. The first hugely contentious issue facing the Apps is the deployment framework, i.e. centralised or decentralised. Based on this, the debate branches out to the corresponding technologies that underpin these architectures, i.e. GPS, QR codes, and Bluetooth. This work conducts a pioneering review of the above scenarios and contributes a geolocation mapping of the current deployment. The Apps' vulnerabilities and the directions of research are identified, with a special focus on the Bluetooth-inspired decentralised paradigm.

**Keywords:** contact tracing app; coronavirus; COVID-19; decentralised; health informatics; mHealth.

**1. Introduction**

In the recent few months, contact-tracing Apps have emerged and pushed the boundary of innovations in response to the outbreak of the coronavirus (COVID-19) [1]. A contact-tracing App [2] is a mobile platform that assists the identification of people who may have come into contact with an infected person, and the subsequent collection of further information about these contacts for containing the virus' spreading. There is an ongoing debate on the deployment of the Apps regarding their technology framework, i.e. centralised [3,4] versus decentralised [5,6], and their corresponding sensor technologies, i.e. the Global Positioning System (GPS) integrated with Quick Response (QR) codes scanning [7,8] and big data analysis [9,10], versus the wireless Bluetooth devices [11] enabled by microwave [12,13] and millimetre-wave [14–17] communications.

In the centralised architecture, personal data collected through the App is controlled by government authority. These Apps mainly follow the PEPP-PT (Pan-European Privacy-Preserving Proximity Tracing) [18] protocol, but the consensus amongst the technical community is that this framework is too academic for practical development. For the decentralized approach, the personal data is enclosed or controlled by individuals only on personal devices. These Apps follow the DP-3T (Decentralised Privacy-Preserving Proximity Tracing) [19] data protection solution recently developed by the European Academics. However, this framework is only partially decentralised, i.e. there is an anonymous centralised database for only the infected people. Google and Apple in partnership [20] will launch an exclusive decentralised framework in May which will be more compatible with the Android and iOS systems. Regarding the technologies and infrastructures that underpin the two architectures, GPS is based on crowd mapping for tracking the spread of the COVID-19, while the QR codes scanning approach is combined with physical temperature testing equipment or thermal imaging cameras to track the healthy or infected individuals' movement on public transport. The Bluetooth method detects other devices retained for a certain amount of time within a certain range of distance, and notifies the devices which have had sufficient contacts with the infected individual's device, assuming that the infected individuals report their anonymous infection states to the App. Researchers from Oxford [21] recently modelled and proposed a threshold on the active user rates (at least 60%) for the App to fully deliver its valuable insights for the government to contain the virus. There is arguably a growing trend globally and especially in Europe that the decentralised architecture would be preferable.

## 2. Survey of the Data Regulations and Technology Protocols

Keeping personal data safe and secure is one of the greatest challenges posed by the rapid development of today's health informatics. The up-to-date regulations and frameworks are detailed in sections below, including the General Data Protection Regulation (GDPR) [22,23], as well as the key competing architectures that have been mentioned in section 1.

*2.1. Data Regulations*

1. GDPR [22,23] released on 14 April 2016 as a standard for Apps - provides the strongest safeguards of trustworthiness (i.e. voluntary approach, data minimisation, time limitation) for Apps to operate widely and accurately to protect personal data and limit intrusiveness.
2. EU approach for efficient contact tracing apps to support gradual lifting of confinement measures [24] released on 15 April 2020 - a common approach for voluntary and privacy-compliant tracing Apps.
3. EU guidelines on the use of location data and contact-tracing tools in the context of the COVID-19 outbreak [25] released on 21 April 2020 - published by European Data Protection Board (EDPB), providing guidance for the use of location data and contact-tracing tools.

*2.2. Technology Protocols*

4. Pan-European Privacy-Preserving Proximity Tracing (PEPP-PT) [26] released on 1 April 2020 – which was followed by the German and Italian governments, and was involved with the development of the UK government's NHSX App (centralised).
5. Decentralized Privacy-Preserving Proximity Tracing (DP-PPT)/ (DP-3T) [27] released on 6 April 2020 - no pooled data is collected, which largely mitigates the privacy risk. The none-infected individuals' data are decentralised based, and the infected individuals' information will be collected anonymously to a central database.
6. Apple and Google partner on COVID-19 contact tracing technology framework [20] (yet to be released in May) - privacy-preserving contact tracing, Bluetooth based, decentralised, free of GPS. Apple and Google tech is currently trading (integrating) with some of the Governments self-running Apps.
7. Government-run contact tracing technology [28] framework that not going to deploy Apple & Google's framework, e.g. the UK, France, and several US states.

## 3. Systematic Mapping of the Global Deployment Status for the COVID-19 Contact-tracing Apps

We produce the first geolocation mapping for the global deployment of the COVID-19 contact-tracing apps in Fig. 1, with the format codes in an order of the country name, App name, the number of users (download times), and the underpinning technologies (GPS, QR codes, Bluetooth). The color of the country represents the employed framework, i.e. with red denoting the centralised architecture, while green representing the decentralised (or being migrating into the decentralised framework, e.g. Austria, Swiss, Estonia, Finland, Germany, Alberta of Canada, and Vietnam).

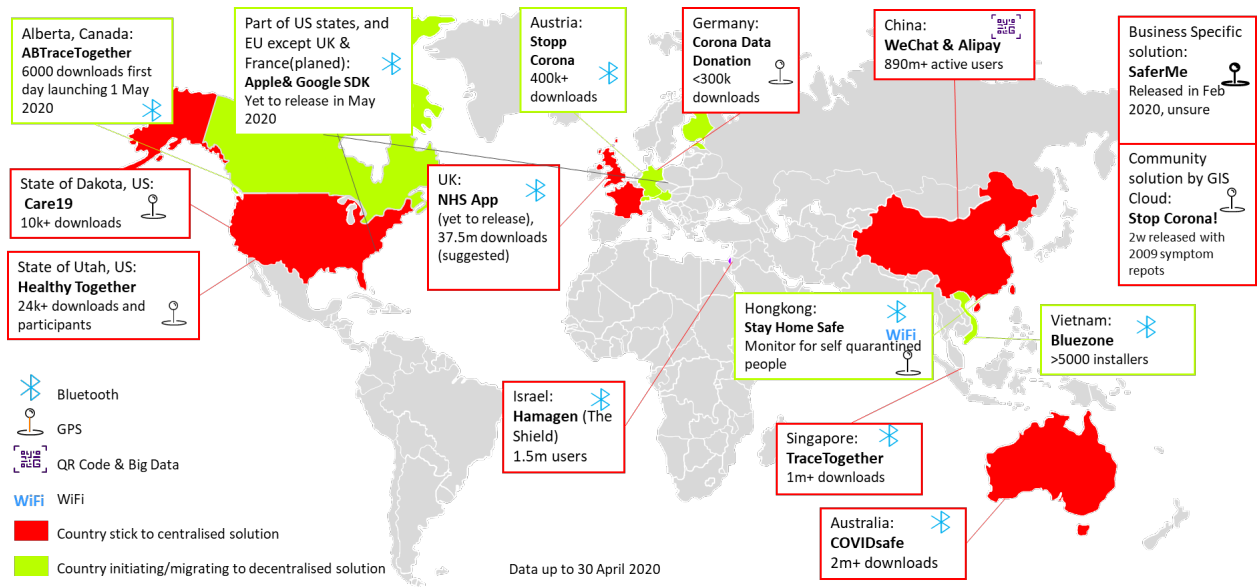

**Figure 1.** Systematic mapping study into the global deployment of the contact-tracing Apps for COVID-19 (format codes order: country name, App/SDK name, download times/user numbers).

Following a software vulnerability-mapping analysis paradigm [24], the flaws of 10 contact-tracing Apps are analysed and summarised below. For instance, one of the key non-technical but important questions for the UK NHS COVID-19 tracking App is the fault positives (i.e. what if people without concern for COVID-19 maliciously report a positive using the App) and the fault negatives (i.e. what if infected people do not report their cases in the App).

(1) The Health Code on Alipay and WeChat (QR code and big data based, centralised) used in China has achieved a 63% of the population coverage and 100% on travelers. However, it is introducing a significant cost for the temperature testing equipment. Many staff are involved in the checking house by house and helping people without using the digital App.

(2) The StayHomeSafe used by Hongkong, CN (Bluetooth, GPS and WiFi based, decentralised) is applied to only self-quarantined people staying at home, and hence not a rigorous contact-tracing App. It is susceptible to the risk if people take off the wristband and phone together and go out.

(3) The Stopp Corona (Bluetooth based, decentralised) used by the Austria Red cross (initiating in Netherlands) is on an ID rolling basis (i.e. public key and private key are rolling) with users covering 4.5% of the population. The Bluetooth's signal-distance relationship and the associated risk level definition are calling for a more unified technical framework to be addressed.

(4) The NHS CV19 App (Bluetooth based) to be deployed by the UK NHS is struggling to address false positives and negatives, especially for the situation if people who have symptom but do not report in the App. There is a rising concern about the App's downloading rates among the whole country, as evidenced in the first launch in the Isle of Wight, with a limited coverage up to 24% of the residents.

(5) The Healthy Together (GPS based, centralised) used by the State of Utah (US), and the Care19 (GPS based, centralised) employed by the North and South Dakota (US) are both vulnerable to flaws in the centralised data pool protection and the sensitive GPS location data issues. The less than 2% of the population coverage also struggles to meet the sensitivity target.

(6) The TraceTogether (Bluetooth based, centralised) proposed by the Singapore Government exhibits limitations in the life or death situations. 16.7% of the population are currently active on the App.
(7) The COVIDsafe (Bluetooth based, centralised) employed by the Australian Department of Health from this May is also susceptible to vulnerabilities in the centralised data pool protection. Currently only 10% of population coverage is reported.
(8) The Hamagen (the Shield) employed by the Israel Health Ministry (Bluetooth based, centralised) reports a user coverage rate of 16.8% and vulnerable to the flaws in the centralised data pool protection.
(9) The BlueZone (Bluetooth based, decentralised) employed by Vietnam provides limited analytics insights for the government, with a user coverage rate less than 0.1%.
(10) The Corona Data Donation (GPS based, centralised) used by the German public health authority is susceptible to the sensitive personal health and location data security problems.

## 4. Challenges and Research Directions for Bluetooth based Contact Tracking Technologies

As observed from Fig.1 and the flaws analysis, Bluetooth (either in centralised or decentralised framework) has accounted for 57% of all the tracking technologies, as compared with the GPS (43%), which merits a further analysis into both the technical and geopolitical characteristics.

Firstly, there is a trade-off between the data privacy and the insights. Arguably, the decentralised and no GPS solution provides the highest level of data protection for individuals as no personal data is collected unless the individual is infected. Without the GPS tracking, Apps cannot collect and trace the movement of the population geographically. With a decentralised framework, however, any data collected from individuals cannot be driven into a centralised database for future analysis, i.e. less information will be provided to the government for controlling the self-quarantine and the movement of the disease among the population.

Secondly, existing decentralised tracing Apps such as the Austria's Stopp Corona are issuing a static unique digital ID to each user with rolling public and private keys (keeping the message encrypted and increasing the data protection standard). If the digital ID is unique and static, it runs the risk that certain digital ID could be hacked and paired with a mobile device, thus compromising the individual privacy. Thereby, a rolling base digital ID to mitigate this vulnerability would be a better practice. In the practical situation, this would be relatively easy to tailor and optimise compared with other related challenges.

Furthermore, different mobile devices exhibit a variety of Bluetooth signal intensity at the ISM band, i.e. the capability of each mobile device to determine the social distance precisely can vary. Accordingly, it is of research and development interest regarding how this can be manipulated (converted) in a unified framework that regulates different generations of devices to communicate and share data with each other. Other factors, such as the multipath interference and spatial blockage between devices are also urgent yet promising research areas that could tip the balance on the functional performance and fault tolerance of the Bluetooth based contact tracking. Coupled with the technical hurdles, the risk-level evaluating standard based on the distance and time contained should be updated accordingly.

## 5. Conclusion

This work reviews the states-of-the-art contact-tracing Apps for the COVID-19. A systematic mapping of the global deployment architectures and technologies is proposed, with a detailed analysis of the flaws for each scenario presented. Specifically, the key challenges facing the Bluetooth based solutions are identified to assist the health informatics decision-making concerning the UK's current status in COVID-19 (see Appendix A for an exponential fitting performed to model the cumulative cases up to date).

**Author Contributions:** Conceptualization, investigation, writing—original draft preparation, X.G. and J.L.; writing—review and editing, J.L.

**Funding:** This research received no external funding.

**Conflicts of Interest:** The authors declare no conflict of interest.

## Appendix A

Fig. 2 below presents an exponential fitting of the cumulative number of COVID-19 cases in the UK from 31 December 2019 to 5 May 2020, with an R square of 97.1%.

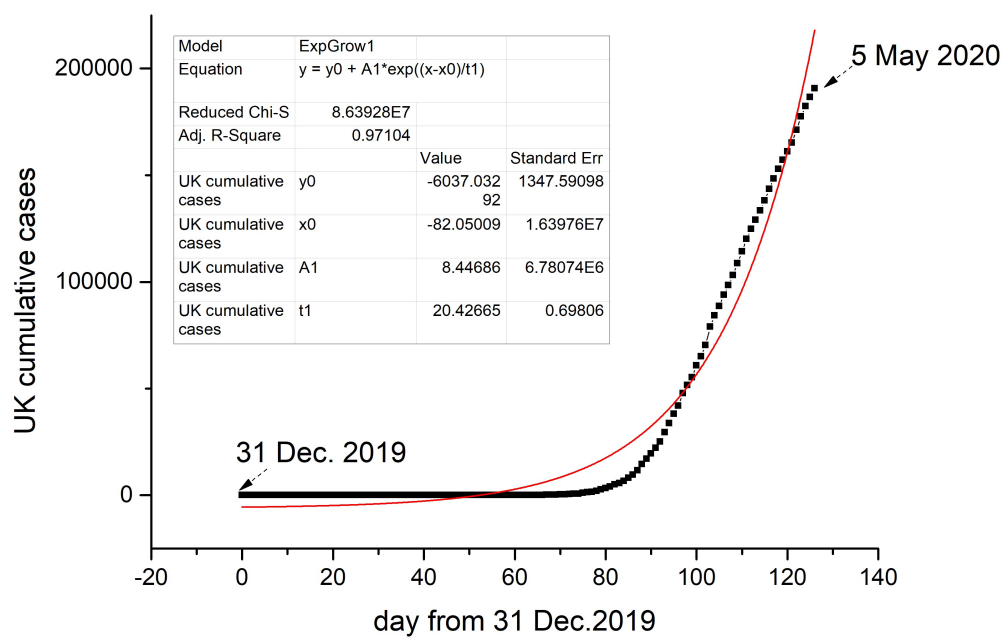

**Figure 2.** Exponential fitting of the cumulative COVID-19 cases in the UK from 31 December 2019 to 5 May 2020.